\documentclass[a4paper]{jpconf}
\usepackage{graphicx}
\begin{document}
\title{Spectral analysis for the {\sc Majorana} {\sc Demonstrator} experiment}

\author{
L Hehn$^1$,
N Abgrall$^1$,
S I Alvis$^2$,
I J Arnquist$^3$,
F T Avignone~III$^{4,5}$,
A S Barabash$^6$,
C J Barton$^7$,
F E Bertrand$^5$,
T Bode$^8$,
A W Bradley$^1$,
V Brudanin$^9$,
M Busch$^{10,11}$,
M Buuck$^2$,
T S Caldwell$^{12,11}$,
Y-D Chan$^1$,
C D Christofferson$^{13}$,
P-H Chu$^{14}$,
C Cuesta$^{2,}$\footnote[23]{Present address:
Centro de Investigaciones Energ\'{e}ticas, Medioambientales y Tecnol\'{o}gicas, CIEMAT, 28040, Madrid, Spain}
J A Detwiler$^2$,
C Dunagan$^{13}$,
Yu Efremenko$^{15,5}$,
H Ejiri$^{16}$,
S R Elliott$^{14}$,
T Gilliss$^{12,11}$,
G K Giovanetti$^{17}$,
M P Green$^{18,11,5}$,
J Gruszko$^2$,
I S Guinn$^2$,
V E Guiseppe$^4$,
C R Haufe$^{12,11}$,
R Henning$^{12,11}$,
E W Hoppe$^3$,
M A Howe$^{12,11}$,
K J Keeter$^{19}$,
M F Kidd$^{20}$,
S I Konovalov$^6$,
R T Kouzes$^3$,
A M Lopez$^{15}$,
R D Martin$^{21}$,
R Massarczyk$^{14}$,
S J Meijer$^{12,11}$,
S Mertens$^{8,22}$,
J Myslik$^1$,
C O'Shaughnessy$^{12,11}$,
G Othman$^{12,11}$,
W Pettus$^2$,
A W P Poon$^1$,
D C Radford$^5$,
J Rager$^{12,11}$,
A L Reine$^{12,11}$,
K Rielage$^{14}$,
R G H Robertson$^2$,
N W Ruof$^2$,
B Shanks$^{12,11}$,
M Shirchenko$^9$,
A M Suriano$^{13}$,
D Tedeschi$^4$,
J E Trimble$^{12,11}$,
R L Varner$^5$,
S Vasilyev$^9$,
K Vetter$^1$,
K Vorren$^{12,11}$,
B R White$^{14}$,
J F Wilkerson$^{12,11,5}$,
C Wiseman$^4$,
W Xu$^7$,
E Yakushev$^9$,
C-H Yu$^5$,
V Yumatov$^6$,
I Zhitnikov$^9$
and
B X Zhu$^{14}$
}

\address{$^1$ Nuclear Science Division, Lawrence Berkeley National Laboratory, Berkeley, CA, USA}
\address{$^2$ Center for Experimental Nuclear Physics and Astrophysics, and Department of Physics, University of Washington, Seattle, WA, USA}
\address{$^3$ Pacific Northwest National Laboratory, Richland, WA, USA}
\address{$^4$ Department of Physics and Astronomy, University of South Carolina, Columbia, SC, USA}
\address{$^5$ Oak Ridge National Laboratory, Oak Ridge, TN, USA}
\address{$^6$ National Research Center ``Kurchatov Institute'' Institute for Theoretical and Experimental Physics, Moscow, Russia}
\address{$^7$ Department of Physics, University of South Dakota, Vermillion, SD, USA} 
\address{$^8$ Max-Planck-Institut f\"{u}r Physik, M\"{u}nchen, Germany}
\address{$^9$ Joint Institute for Nuclear Research, Dubna, Russia}
\address{$^{10}$ Department of Physics, University, Durham, NC, USA}
\address{$^{11}$ Triangle Universities Nuclear Laboratory, Durham, NC, USA}
\address{$^{12}$ Department of Physics and Astronomy, University of North Carolina, Chapel Hill, NC, USA}
\address{$^{13}$ South Dakota School of Mines and Technology, Rapid City, SD, USA}
\address{$^{14}$ Los Alamos National Laboratory, Los Alamos, NM, USA}
\address{$^{15}$ Department of Physics and Astronomy, University of Tennessee, Knoxville, TN, USA}
\address{$^{16}$ Research Center for Nuclear Physics, Osaka University, Ibaraki, Osaka, Japan}
\address{$^{17}$ Department of Physics, Princeton University, Princeton, NJ, USA}
\address{$^{18}$ Department of Physics, North Carolina State University, Raleigh, NC, USA}	
\address{$^{19}$ Department of Physics, Black Hills State University, Spearfish, SD, USA}
\address{$^{20}$ Tennessee Tech University, Cookeville, TN, USA}
\address{$^{21}$ Department of Physics, Engineering Physics and Astronomy, Queen's University, Kingston, ON, Canada} 
\address{$^{22}$ Physik Department, Technische Universit\"{a}t, M\"{u}nchen, Germany}

\ead{lhehn@lbl.gov}

\begin{abstract}
The {\sc Majorana} {\sc Demonstrator} is an experiment constructed to search for neutrinoless double-beta decays in germanium-76 and to demonstrate the feasibility to deploy a ton-scale experiment in a phased and modular fashion. It consists of two modular arrays of natural and $^{76}\textrm{Ge}$-enriched germanium detectors totaling 44.1\,kg (29.7\,kg enriched detectors), located at the 4850' level of the Sanford Underground Research Facility in Lead, South Dakota, USA. Data taken with this setup since summer 2015 at different construction stages of the experiment show a clear reduction of the observed background index around the ROI for $0\nu\beta\beta$-decay search due to improvements in shielding. We discuss the statistical approaches to search for a $0\nu\beta\beta$-signal and derive the physics sensitivity for an expected exposure of $10\,\textrm{kg}{\cdot}\textrm{y}$ from enriched detectors using a profile likelihood based hypothesis test in combination with toy Monte Carlo data.
\end{abstract}

\section{Introduction}
The {\sc Majorana} {\sc Demonstrator} is a rare event search experiment designed to search for the neutrinoless double beta decay ($0\nu\beta\beta$) of $^{76}\textrm{Ge}$~\cite{mjd}. A signal for this yet unobserved decay mode would manifest itself as a sharp peak at the Q-value of 2039\,keV of the observed, continuous double beta decay spectrum of $^{76}\textrm{Ge}$. In order to reduce the background due to atmospheric cosmic muons, the experiment is located at the 4850' level of the Sanford Underground Research Facility (SURF) in Lead, South Dakota. There, the experiment is housed in a clean room of class 1000. Both source and detectors for possible $0\nu\beta\beta$ electrons are high-purity Germanium (HPGe) crystals operated at liquid nitrogen temperature in two separate cryostats made of ultra-clean, electroformed Cu. To allow for a naturally scalable setup each cryostat is combined with a cooling system and electronics into an independent module. Background from internal components is significantly reduced through the selection of radiopure materials~\cite{mjdassay}. The detectors are additionally shielded from ambient radiation by several layers of low-background passive shielding: an inner and outer copper shield, a lead shield, and a radon enclosure. The whole experiment is furthermore surrounded by the panels of an active muon veto to tag atmospheric muons followed by a thick outermost layer of polyethylene.

In its current phase, detectors with a combined mass of 44.1\,kg are installed. Of this mass, 29.7\,kg comprise detectors with an enrichment fraction of $^{76}\textrm{Ge}$ of 88\%. The remaining 14.4\,kg are detectors made from Germanium with a natural abundance. All enriched detectors are of P-type Point Contact (PPC) design. With their low capacitance and optimized read-out electronics they offer low noise and superb energy resolution of ${\approx}0.1\%$ at $Q_{\beta\beta}$. In addition, their electric field configuration allows for efficient pulse-shape analysis to reject backgrounds, including those from multi-site interactions within the detector~\cite{clara}, and alpha decays on the detector surface~\cite{dcr1,dcr2}.

While the experiment is optimized for the search for $0\nu\beta\beta$, the performance of the detectors, e.g.~low energy threshold and background, allows for other searches for physics beyond the standard model~\cite{mjdaxions}. The low background level achieved with the {\sc Majorana} {\sc Demonstrator} is compatible~\cite{vince} with the design goal of ${<}3\,\textrm{cnts/ROI/ton/y}$.

\section{\label{sec:statisticalmethods}Statistical methods for physics searches}
A variety of Bayesian and frequentist statistical methods are in use by the experiments searching for neutrinoless double beta decay. In~\cite{nu2016}, the {\sc Majorana} Collaboration used the Feldman-Cousins method~\cite{feldmancousins} on a first dataset of the {\sc Majorana} {\sc Demonstrator} to derive an upper limit on the $0\nu\beta\beta$ half-life. The F.-C.-method eliminates the so-called flip-flop problem, automatically transitioning the confidence interval from an upper limit to a two-sided confidence interval. However, this method is somewhat conservative and the reported confidence interval can exhibit a significant amount of over-coverage leading to a reduced sensitivity for exclusion limits~\cite{overcoverage}.
 
A different frequentist method which can mitigate coverage problems to some degree is based on the profile likelihood. Further advantages of this method are the simple implementation of systematic uncertainties via constraints on nuisance parameters. By multiplying likelihood functions it is possible to analyze multiple datasets simultaneously using a common parameter of interest. When dealing with likelihood functions for a large number of events---the so-called large-sample case---one can apply Wilks' theorem~\cite{wilks}. It approximates the shape of the profile likelihood function allowing to construct either one- or two-sided confidence intervals with reduced computational effort (see e.g.~\cite{exo}). For smaller samples $\mathcal{O}(10)$ events the approximation does not hold and hypothesis tests using Monte Carlo generated toy datasets are required (see e.g.~\cite{gerda}). A modification of this method, often referred to as {\it ``CLs''}-method~\cite{cls}, can be used to compensate for underfluctuations of the background (see e.g.~\cite{nemo}). In the following we focus on the profile likelihood method in the small sample case and use toy Monte Carlo data to derive an upper limit. Results based on the CLs-method are however given for comparison.

\section{Likelihood model validation and projected sensitivity}
The likelihood function used in this analysis is identical to the simple model which was employed in~\cite{gerda} by the GERDA collaboration. For a single dataset $\mathcal{D}$ of $N^\textrm{obs}$ event energies $E_\mathrm{n}$, the probability density function (PDF) is constructed from a Gaussian shaped signal and a flat background in the considered energy range ${\Delta E}$:

\begin{equation}
\label{equationlikelihood}
	\mathcal{L} (\mathcal{D}|{\cal S}, \textrm{BI}, \theta) \ =\
	\prod_{n=1}^{N^\textrm{obs}}  \   
	\frac{1}{\mu^{S} + \mu^{B}} \, \cdot\\ 
	\left[  
		\mu^{S} \cdot \frac{1}{\sqrt{2\pi} \sigma}
		\exp\left(\frac{-(E_\mathrm{n}-Q_{\beta\beta} - \delta)^2}{2 \sigma^2}\right) +
		\mu^{B} \cdot \frac{1}{\Delta E}
	\right].
\end{equation}

\noindent The rate of signal events $\mu^{S}  = \ln 2 \cdot \left(N_{A}/m_{a}\right) \cdot \epsilon \cdot {\cal E} \cdot {\cal S}$ depends on the inverse half-life ${\cal S}$ (the parameter of interest), the raw detector exposure ${\cal E}$, and the efficiency $\epsilon$ to detect $0\nu\beta\beta$-events ($N_{A}$ is the Avogadro constant and $m_{a}$ the molar mass of $^{76}\textrm{Ge}$). The rate of background events $\mu^{B}  = {\cal E} \cdot \textrm{BI} \cdot \Delta E$ is proportional to the background index $\textrm{BI}$. The efficiency $\epsilon$, the width $\sigma$ and the possible shift $\delta$ of the Gaussian signal are considered nuisance parameters $\theta = \{\epsilon, \sigma, \delta\}$ and constrained in the fit. The likelihood functions for multiple datasets are multiplied to perform a simultaneous fit with a common ${\cal S}$. 

A limit on ${\cal S}$ can be found by calculating the profile likelihood test statistic 

\begin{equation}
t_{\cal S} = 
-2 \ln {\lambda}({\cal S}) = -2 \ln \frac
{\mathcal{L} ( {\cal S},\hat{\hat{\bf BI}},\hat{\hat{\theta}} )}
{\mathcal{L} ( \hat{\cal S},\hat{\bf BI},\hat{\theta} )}
\end{equation}

\noindent for toy datasets generated under the assumption of different values ${\cal S}_\textrm{j}$. The value of ${\cal S}$ which can be excluded at 90\% C.L. is found when the p-value of the observed data, $p_{{\cal S}_\textrm{j}} = \int_{t_\textrm{obs}}^\infty  f(t_{\cal S}|{\cal S}_\textrm{j}) \, d({t_{{\cal S}}}_\textrm{j})$, with respect to the toy data distribution is $\leq 0.1$. The median sensitivity is determined similarly from the median of the distribution from toy datasets generated under the background-only hypothesis.

\begin{figure}[h]
\begin{minipage}{17pc}
\includegraphics[width=17pc]{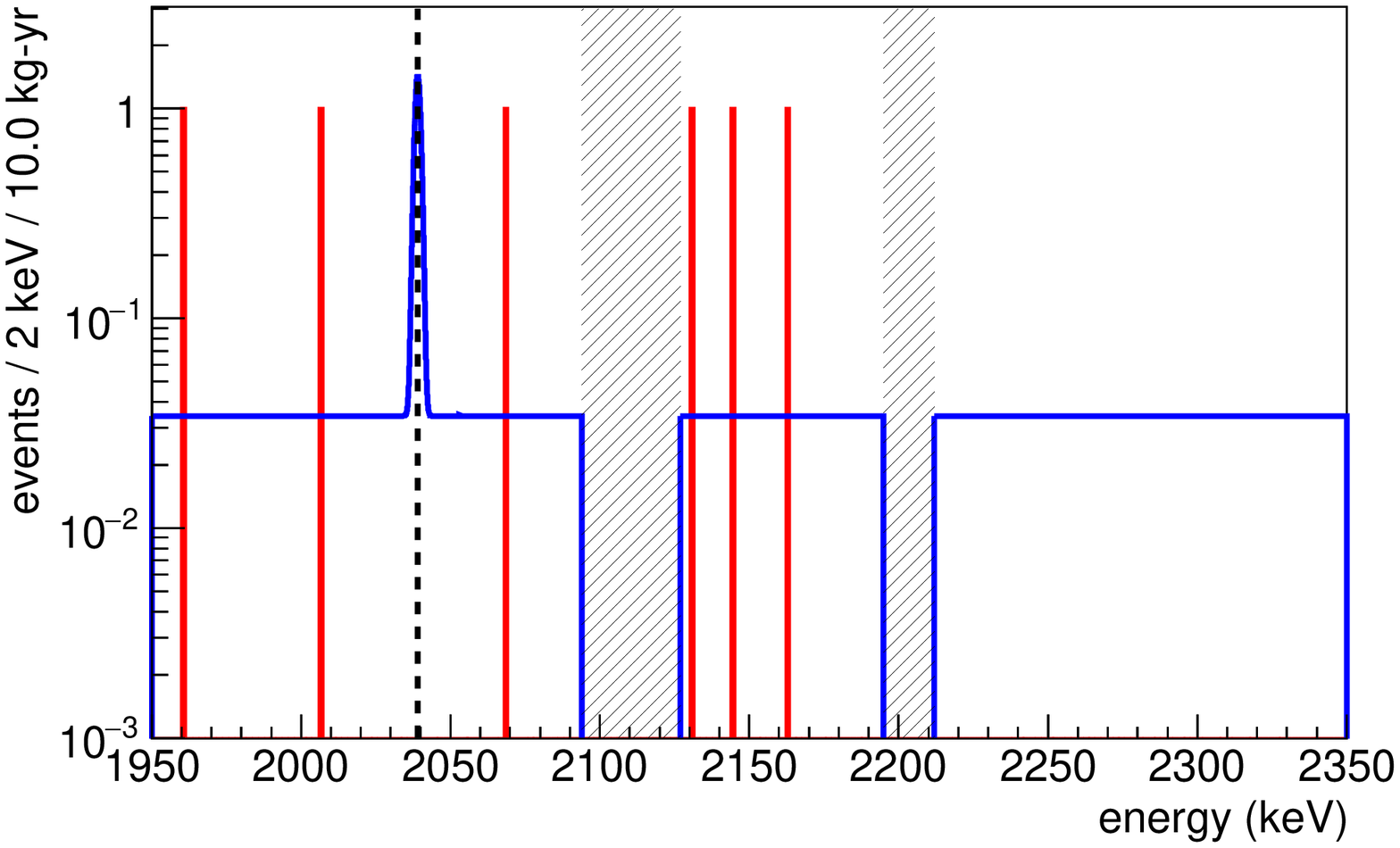}
\caption{\label{fig:bestfit}B-only toy dataset (red histogram) for $10\,\textrm{kg}{\cdot}\textrm{y}$ of {\sc Majorana} {\sc Demonstrator} data with best fit background {+}68\% upper error on signal (blue). Grey shaded energy regions are excluded from the fit due to expected background.\newline\newline\newline}
\end{minipage}\hspace{2pc}%
\begin{minipage}{17pc}
\includegraphics[width=17pc]{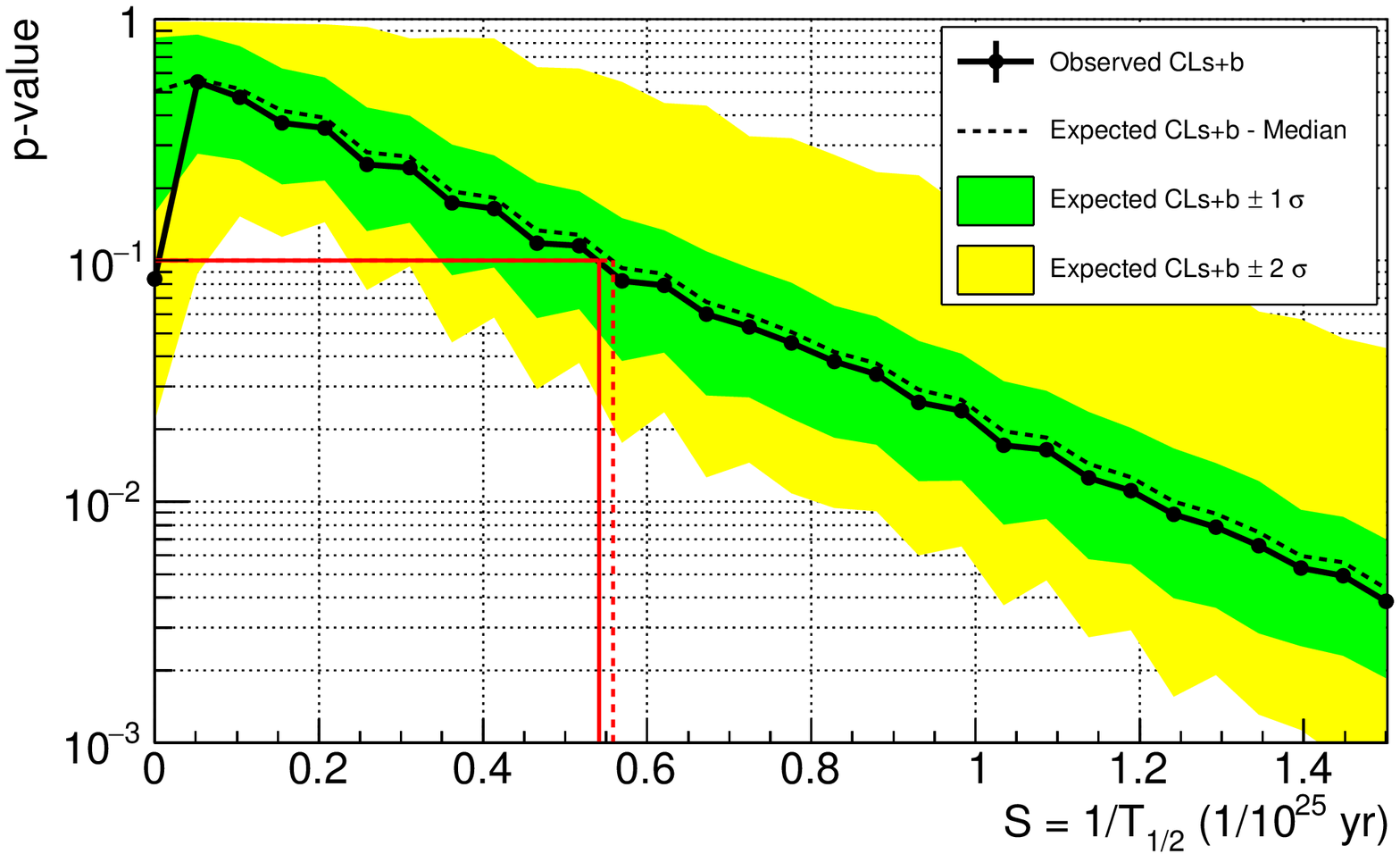}
\caption{\label{fig:hyposcan}Resulting p-values from hypothesis tests for a range of inverse half-life values ${\cal S}_\textrm{j}$ (black markers). The lowest value of $\cal S$ excluded at 90\% C.L. for the toy dataset is denoted with a solid red line. The dashed black line and colored bands are the median sensitivity with $1\,\sigma$ and $2\,\sigma$ uncertainty derived from B-only toy data.}
\end{minipage} 
\end{figure}

The implementation of this likelihood model and the statistical hypothesis tests were performed with the \texttt{RooStats} framework~\cite{roostats}. A validation of the code based on this framework was performed with data and parameters from~\cite{gerda}. The information on parameter values and background levels provided in~\cite{gerda} allowed to partially reconstruct the likelihood model and derive the median sensitivity of the GERDA experiment of $T_{1/2} \geq 4.0 \cdot 10^{25}\,\textrm{y}$ to an accuracy of 2\%.

\begin{table}
\caption{\label{table}Parameter values and constraint uncertainties for the single toy dataset used to determine the {\sc Majorana} {\sc Demonstrator} sensitivity.}
\begin{center}
\begin{tabular}{l c}
\br
model parameter & value\\
\mr
$\textrm{BI}$ (cnts/keV/kg/y) & $1.8 \cdot 10^{-3}$ \\
$\cal E$ ($\textrm{kg} \cdot \textrm{y}$) & $10.0 \pm 0.2\phantom{0}$ \\
$\sigma$ (keV) & $1.00 \pm 0.01$ \\
$\epsilon$ & $0.590 \pm 0.059$ \\
$\delta$ (keV) & $0.0 \pm 0.2$ \\
\br
\end{tabular}
\end{center}
\end{table}

A full statistical analysis of open data from the {\sc Majorana} {\sc Demonstrator} is in progress. Demonstrated here is the application of this likelihood method on a set of toy data to determine the sensitivity of the experiment. A single toy dataset was generated under the background-only hypothesis, with a background index as measured for datasets 3 and 4 and an exposure similar to the combination of all datasets. Typical values representative of the available {\sc Majorana} {\sc Demonstrator} datasets were chosen for all parameters in equation~\ref{equationlikelihood} and are listed in table~\ref{table}. The energy range considered goes from 1950 to 2350\,keV, where the regions 2094\,keV to 2127\,keV and 2195\,keV to 2212\,keV have been excluded due to $\gamma$-peaks expected from the background model. In this 350\,keV energy range, the generated toy dataset  contains 6 events. Figure~\ref{fig:bestfit} shows a fit of this toy dataset resulting in a best fit of zero signal events. All nuisance parameters are fitted with their expected constraint values. Using 100{,}000 toy Monte Carlo datasets generated for each of the ${\cal S}_\textrm{j}$+B hypotheses and 20{,}000 for the B-only hypotheses we derive the median sensitivity for the {\sc Majorana} {\sc Demonstrator} based on the previously observed background index quoted in table~\ref{table}. As shown in Figure~\ref{fig:hyposcan} for this toy dataset, the $0\nu\beta\beta$ half-life that can be excluded at 90\% C.L. is $T_{1/2} \geq 1.8 \cdot 10^{25}\,\textrm{y}$ , which given the exposure corresponds to $\mu^{S} \geq 1.8$ signal events. The sensitivity achieved with this method is therefore significantly higher than what we derive with the Feldman-Cousins method, which excludes approximately $\mu^{S} \geq 2.4$ signal events. Based on the same toy Monte Carlo datasets exclusion limits using the CLs-method are also computed, resulting in a more conservative half-life limit of $T_{1/2} \geq 1.3 \cdot 10^{25}\,\textrm{y}$, corresponding to $\mu^{S} \geq 2.5$ signal events.

\section{Outlook}
While the sensitivity study presented here is based on a single toy dataset with an exposure of $10\,\textrm{kg}{\cdot}\textrm{y}$, a full analysis of open data from the {\sc Majorana} {\sc Demonstrator} is currently ongoing and the results are expected to be published soon. For the profile likelihood based analysis, data from all available datasets 0 to 5 is used in a simultaneous fit, while taking into account different characteristics of the signal shape for individual datasets. In parallel, the {\sc Majorana} collaboration is complementing these results employing several of the statistical methods described in section~\ref{sec:statisticalmethods}, including Bayesian methods, to search for new physics such as $0\nu\beta\beta$. 
 
Since October 2016, data taking with the fully shielded setup and the readout of both detector modules in an integrated DAQ mode is ongoing. For this additional data a blinding scheme is in effect. With the achieved low background levels and running time of 3-5 years the {\sc Majorana} {\sc Demonstrator} is aiming for a projected sensitivity of $T_{1/2} \geq 10^{26}\,\textrm{y}$ at 90\% C.L.

\ack
This material is based upon work supported by the U.S. Department of Energy, Office of Science, Office of Nuclear Physics, the Particle Astrophysics and Nuclear Physics Programs of the National Science Foundation, and the Sanford
Underground Research Facility.

\end{document}